\documentclass{article}

\usepackage{iclr2027_conference,times}

\usepackage{amsmath,amsfonts,bm}

\def\eqref#1{equation~\ref{#1}}

\def\1{\bm{1}}

\DeclareMathAlphabet{\mathsfit}{\encodingdefault}{\sfdefault}{m}{sl}
\SetMathAlphabet{\mathsfit}{bold}{\encodingdefault}{\sfdefault}{bx}{n}

\usepackage[utf8]{inputenc}
\usepackage[T1]{fontenc}

\usepackage{hyperref}
\usepackage{url}

\usepackage{booktabs}
\usepackage{amsfonts}
\usepackage{amsmath}
\usepackage{amssymb}
\usepackage{nicefrac}
\usepackage{microtype}
\usepackage{xcolor}

\usepackage{graphicx}
\usepackage{subcaption}
\usepackage{tabularx}
\usepackage{multirow}
\usepackage{wrapfig}

\usepackage{algorithm}
\usepackage{algpseudocode}
\usepackage{float}

\usepackage{enumitem}
\usepackage{xspace}

\newcommand{\sys}{\textit{GoodServe}\xspace}

\newcommand{\phm}[1]{%
    \vspace{.4em}
    \noindent\textbf{#1}\hspace{.5em}
}

\title{
\emph{GoodServe}: Globally-coordinated Routing of Agentic
LLM Inferences for High Goodput
}

\author{%
\textbf{Boxiao Du}\textsuperscript{1,*} \quad
\textbf{Boning Huangfu}\textsuperscript{1,*} \quad
\textbf{Yizhou Luo}\textsuperscript{1} \quad
\textbf{Chen Chen}\textsuperscript{1,\textdagger} \\[0.3em]
\textbf{Zijun Li}\textsuperscript{1} \quad
\textbf{Minchen Yu}\textsuperscript{2} \quad
\textbf{Xiaoyi Fan}\textsuperscript{3} \quad
\textbf{Minyi Guo}\textsuperscript{1} \\[0.8em]
\normalfont
\textsuperscript{1}Shanghai Jiao Tong University \\
\textsuperscript{2}The Chinese University of Hong Kong, Shenzhen \\
\textsuperscript{3}Shenzhen MSU-BIT University \\[0.4em]
\normalsize
\textsuperscript{*}These authors contributed equally to this work. \\
\textsuperscript{\textdagger}Corresponding author.%
}

\iclrfinalcopy

\begin{document}

\maketitle

\fancyhead[L]{Preprint}
\fancyhead[C]{}
\fancyhead[R]{}

\begin{abstract}
Agentic applications consume LLM inferences as complete outputs: a code patch or a database query is useless until its last token arrives, so what matters to the application is end-to-end latency against a deadline, and what matters to the operator is \emph{goodput}, the number of inferences that finish within their deadlines.
Operators usually serve these inferences on pools of heterogeneous GPUs, and every request must be routed to one instance on arrival.
Existing routers are greedy: each request goes to the instance that would serve it fastest or is least loaded, which spends the strongest GPUs on requests that would have met their deadlines on weaker ones and leaves urgent requests waiting.
Goodput instead calls for \emph{globally coordinated} routing, in which a request may be placed sub-optimally for itself so that the pool meets more deadlines overall.
Coordination needs two quantities a greedy router never estimates: how long each request will run, since its output length is unknown until it finishes, and how fast each instance currently is.
\emph{GoodServe} formulates routing as a global goodput problem whose latency model couples both, including prefix-cache reuse, and solves it in a \emph{predict-and-rectify} manner: a mixture-of-experts predictor gated by task type estimates output length accurately and efficiently; an EMA-smoothed black-box estimator tracks each instance's queuing delay and per-token speed; a \emph{just-enough} rule sends each request to the weakest instance that can still meet its deadline and rejects requests that no instance can serve in time; and periodic risk checks migrate endangered requests by token IDs, $7$--$15\times$ cheaper than transferring KV cache.
On a four-GPU heterogeneous testbed serving three agentic benchmarks, \emph{GoodServe} attains the highest goodput at every SLO level, up to 31.0\% over the best existing router, and its routing overhead stays at 5\,ms per request at 10,000 requests per second across 512 simulated instances.
\end{abstract}


\section{Introduction}
\label{sec:intro}

An agentic application does not consume an LLM's answer token by token; it waits for the whole answer and then acts on it.
A coding agent applies a patch only once the patch is complete~\cite{liu2025sew}, a database agent executes a query only once the query is complete~\cite{phan2025askdb}, and a reasoning agent cannot verify a chain of thought before it ends~\cite{yan2025mathagent}.
The latency that matters to such applications is therefore the \emph{end-to-end} latency of an inference, from request to last token, rather than time-to-first-token or time-per-output-token, and users attach explicit end-to-end deadlines (E2E-SLOs) to agentic inferences~\cite{zhang2025jitserve,kim2026kairos,yu2026superinfer,chow2025slice,bari2025optimal}.
What matters to the operator is then \emph{goodput}: the number of inferences completed within their deadlines.

\begin{figure}[t]
  \centering
  \begin{minipage}[t]{0.40\textwidth}
    \centering
    \includegraphics[width=0.78\linewidth]{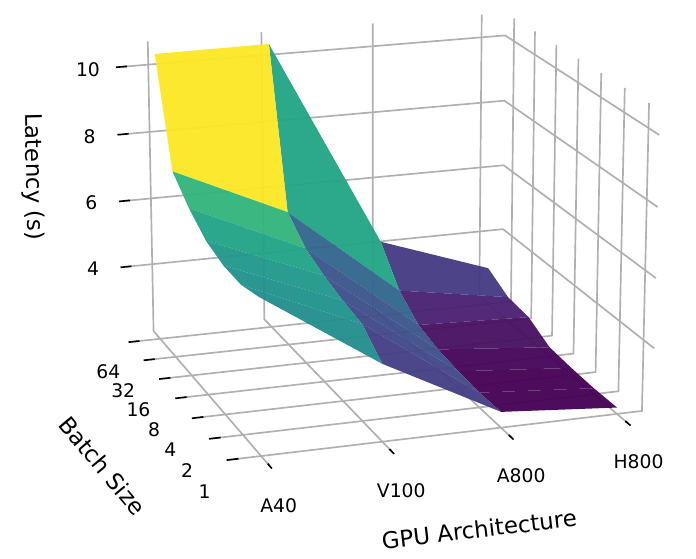}
    \caption{Inference latency across four GPU architectures under varying batch sizes, for a fixed sequence of 100 input and 200 output tokens.}
    \label{fig:gpu-letancy}
  \end{minipage}\hfill
  \begin{minipage}[t]{0.56\textwidth}
    \centering
    \includegraphics[width=0.60\linewidth]{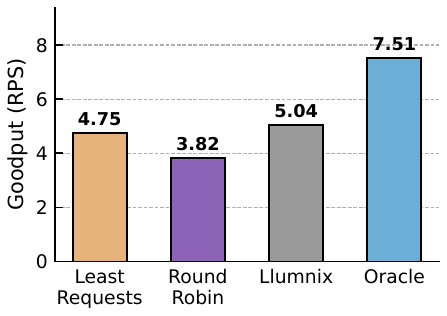}
    \caption{Performance inferiority of existing routing strategies: 600 requests (10 requests per second) served by four heterogeneous GPUs (V100, A40, A800, H800); each request has 100 input tokens and an output length sampled from [100, 500]; the E2E-SLO is 6\,s.}
    \label{fig:gpu-letancy-1}
  \end{minipage}
\end{figure}

Operators serve these inferences on whatever GPUs they own, and what they own is often heterogeneous: GPU generations are purchased over years and pooled together, as in Microsoft's Office~365 Copilot pool of H100, A100, and MI300X GPUs~\cite{asgar2025efficient,mei2025helix}.
Every incoming request is routed by a proxy to one instance in the pool, and that decision fixes how fast the request will be served: the same inference runs several times slower on a V100 than on an H800, and slower again as the batch on an instance grows (Fig.~\ref{fig:gpu-letancy}).
Routing therefore decides goodput.

Existing routers decide it badly, because they are \emph{greedy}.
Random, round-robin, and least-request routing~\cite{rayserve_routing_policies}, lowest-TPM routing~\cite{litellm2026}, Llumnix~\cite{sun2024llumnix}, and prefix-cache-aware routers such as Preble~\cite{srivatsa2024preble} all answer one question per request: which instance serves \emph{this} request fastest, or is least loaded right now.
None of them asks whether the request needs to be served fast at all.
A request with a loose deadline is still sent to the strongest idle GPU, where it occupies capacity that a later, urgent request will need, while a weaker GPU that would have met the loose deadline stays idle.
Goodput requires the opposite discipline, \emph{globally coordinated} routing: place each request only as well as its deadline demands, so that the pool as a whole meets as many deadlines as possible, even if some requests are placed sub-optimally for themselves.
The gap between the two is large: on a four-GPU testbed, an oracle router that knows each request's output length and each GPU's speed reaches a goodput of 7.51 requests per second, whereas least-request, round-robin, and Llumnix reach 4.75, 4.82, and 5.04 (Fig.~\ref{fig:gpu-letancy-1}), a 49--58\% loss that no load balancing recovers.

Coordinated routing is hard for two reasons.
First, it must weigh many factors at once, on both the \emph{demand} and the \emph{resource} side.
Whether a weaker instance can still meet a deadline depends on the request's input and output lengths, on how much of its prompt that instance already holds in its prefix cache, and on that instance's queuing delay and per-token prefill and decode speed; without the resource side, a router cannot tell a safe demotion from a fatal one, which is exactly why greedy routers, having no such estimate, play safe and send everything to the strongest instance.
Second, the most important factor, the output length, is unknown until the request finishes, and instance speeds drift with load, so any estimate will sometimes be wrong; a router that acts on estimates must also be able to correct them.

\sys is a goodput-centric router for agentic LLM inferences over heterogeneous GPUs built on these two observations.
It formulates routing as a global goodput-maximization problem whose latency model couples demand and resource terms, including prefix-cache reuse (Sec.~\ref{sec:problem}), and it solves that problem in a \emph{predict-and-rectify} manner.
For demand, a mixture-of-experts predictor exploits the fact that in agentic workloads the output length is governed by an implicit task type: small experts specialize in length tiers and a gate conditioned on the prompt selects among them, at 2.5\,ms per request.
For resources, an EMA-smoothed black-box estimator learns each instance's queuing delay and per-token speeds from its observed service times, requiring no access to engine internals (Sec.~\ref{sec:estimation}).
For the decision, a \emph{just-enough} rule sends each request to the weakest instance that is still predicted to meet its deadline, which is the per-request realization of the global principle above and costs $O(M)$ over $M$ instances; requests with no feasible instance are rejected, and every $\tau$ iterations \sys re-evaluates the deadline risk of active requests and migrates endangered ones to a stronger instance by transferring their token IDs rather than their KV cache, which is $7.1$--$15.3\times$ faster (Sec.~\ref{sec:routing}).

We implemented \sys in 2.5k lines of code and evaluated it on a testbed of four heterogeneous GPUs (V100, A40, A800, H800) serving Llama-3.1-8B and Qwen2.5-14B on a workload mixed from BIRD-bench, SWE-bench, and LiveCodeBench with production arrival traces.
\sys delivers the highest goodput and lowest SLO violation ratio at every SLO level, up to 31.0\% over the best existing router, and its advantage grows as deadlines loosen, exactly where coordinated routing has room to act.
An oracle output-length predictor adds only 5.1\% more goodput, so the routing principle, not prediction accuracy, carries most of the gain; the gain persists under bursty arrivals, out-of-distribution tasks, and a predictor transferred from another model, and the router's overhead stays at 5\,ms per request at 10,000 requests per second across 512 simulated instances.

\section{Background and Motivation}
\label{sec:background_motivation}

\subsection{Serving Agentic LLM Inferences over Heterogeneous Instances}
\label{sec:bg_llm_serving}

Confronting booming inference demand, production providers maintain a pool of LLM serving instances, and a service proxy routes every arriving request to one instance for execution.
The serving capability of these instances is \emph{inconsistent}, for two reasons.
The first is static hardware heterogeneity: providers keep purchasing new GPUs while older ones remain in service, so the pool spans generations~\cite{mei2025helix,asgar2025efficient}; Microsoft, for example, serves the Office~365 Copilot workload from a cluster mixing NVIDIA H100, A100, and AMD MI300X GPUs~\cite{asgar2025efficient}.
The second is dynamic fluctuation: the per-iteration time of an instance depends on the batch it is currently serving, which changes with the instantaneous request intensity.
Fig.~\ref{fig:gpu-letancy} shows both effects for a fixed sequence across A40, V100, A800, and H800 GPUs and varying batch sizes.

For agentic LLM inferences, whose utility is binary upon completion~\cite{zhang2025jitserve,kim2026kairos} and which carry explicit E2E-SLOs~\cite{yu2026superinfer,zhang2025jitserve,chow2025slice,bari2025optimal,asgar2025efficient}, the routing decision over such a pool determines whether a request meets its deadline, and hence the goodput of the service.
Given the heterogeneity above and the uncertainty of decode lengths, making that decision well is both crucial and challenging.
We next examine how existing routers make it.

\subsection{Lessons Learned from Existing Request-routing Methods}
\label{sec:mot_challenges}

A series of routing methods is used in practice.
\emph{Random} (power of two choices)~\cite{ray_serve_docs} and \emph{round-robin}~\cite{yu2025efficient} spread requests evenly across instances.
\emph{Least-request} and \emph{prefix-cache} routing are built into production frameworks such as AIBrix~\cite{team2025aibrix}: the former routes to the instance with the fewest pending requests, the latter to the instance with the highest cache hit ratio.
\emph{Lowest-TPM} routing in LiteLLM~\cite{litellm2026} picks the instance with the least tokens-per-minute utilization, the \emph{Preble}-style method~\cite{srivatsa2024preble,qin2025mooncake} weighs prefix-cache hits against compute load to find the best-performing instance for a request, and Llumnix~\cite{sun2024llumnix} routes to the instance with the most available memory and migrates requests at run time to rebalance load.

All of these are \emph{greedy} and \emph{local}: each request is placed where it, taken alone, would run fastest or where the load is lightest, without regard to its own deadline or to the requests that will arrive after it.
They therefore treat urgent and relaxed requests alike, and cannot make a \emph{locally-suboptimal yet globally-optimal} decision, such as sending a relaxed request to a weak instance to keep a strong one free.
Fig.~\ref{fig:gpu-letancy-1} quantifies the cost on a four-GPU testbed: an oracle router with ground-truth knowledge of output lengths and instance speeds (policy in Sec.~\ref{sec:routing}) reaches a goodput of 7.51, whereas least-request, round-robin, and Llumnix reach 4.75, 4.82, and 5.04.

The lesson is that goodput requires \emph{globally coordinated}, SLO-adaptive routing that reasons about both the demand of each request (its decode length) and the capability of each resource (its speed and queue), rather than a fixed local heuristic.
Doing so under the uncertainty of both quantities is the problem this paper addresses.

\section{\sys Design}
\label{sec:solution}

\subsection{Goodput-Optimal Routing as a Global Problem}
\label{sec:problem}

We first state the routing problem precisely, because the statement itself carries the first contribution of this paper: goodput-optimal routing is a \emph{global} problem over all requests and all instances, and its latency model must couple the \emph{demand} side of a request with the \emph{resource} side of an instance.
A detailed symbol table is given in Appendix~\ref{subsec:notion}.

Given the request set $\mathcal{R}$ and the GPU instance set $\mathcal{G}$, let $x_{r,g}\in\{0,1\}$ indicate whether request $r$ is routed to GPU-$g$, and let $T(r,g)$ be the end-to-end latency of request-$r$ when served on GPU-$g$.
The objective is to maximize goodput,
\begin{equation}
\max_{\{x_{r,g}\}} \ \sum_{r\in\mathcal{R}} \mathbb{I}\!\left[\sum_{g\in\mathcal{G}} x_{r,g}\,{T}(r,g) \leq D_r\right],
\label{eq:objective}
\end{equation}
where $D_r$ is the E2E-SLO of request-$r$, subject to $\sum_{g} x_{r,g}=1$ for every request and to the memory capacity of each GPU, which bounds the number of requests it can serve concurrently~\cite{vllm_github}.

The latency $T(r,g)$ is where demand meets resource.
On the demand side, let $L^{\mathrm{in}}_r$ and $L^{\mathrm{out}}_r$ be the input and output lengths of request-$r$; on the resource side, let $q_g$ be the average queuing delay on GPU-$g$ and $p_g$ and $d_g$ its average per-token prefill and decode latencies.
Prefix caching, which production systems adopt to skip the prefill of cached prefixes~\cite{qin2025mooncake}, enters through the hit length $H_{r,g}$ of request-$r$ on GPU-$g$, so that
\begin{equation}
{T}(r,g) \triangleq {q}_g + p_g \cdot (L^{\mathrm{in}}_r - H_{r,g}) + d_g \cdot L^{\mathrm{out}}_r.
\label{eq:e2e}
\end{equation}
Eq.~\ref{eq:e2e} makes three things explicit that greedy routers leave implicit.
Routing is coupled across requests, since $q_g$ and the capacity of GPU-$g$ depend on the requests already placed there.
Feasibility on a weaker instance can only be judged with the resource terms, since a large $d_g$ may be tolerable for a short $L^{\mathrm{out}}_r$ and fatal for a long one.
And cache locality is not a separate objective but a latency term: a long cached prefix $H_{r,g}$ lowers $T(r,g)$ and can make a busier or weaker instance feasible, so \sys honors locality exactly to the extent that it helps meet the deadline.

\phm{Why the problem is hard.}
Even with exact coefficients, Problem~\ref{eq:objective} is NP-hard: with binary routing variables and bounded per-GPU capacities, choosing which requests to serve on which GPU so as to satisfy the most deadlines contains the multiple-knapsack problem, where each GPU is a knapsack whose capacity is consumed by the queuing delay and memory that admitted requests add~\cite{zhang2025jitserve}.
The coefficients are, moreover, not exact: $L^{\mathrm{out}}_r$ is unknown until the request finishes, and $q_g$, $p_g$, $d_g$ drift with the load on each instance.
\sys therefore adopts a \emph{predict-and-rectify} principle.
It estimates the demand and resource coefficients accurately enough to make a coordinated decision (Sec.~\ref{sec:estimation}), decides with a rule that realizes the global principle at $O(M)$ cost per request (Sec.~\ref{sec:routing}), and revisits the decision while the request runs so that estimation errors are corrected rather than suffered.

\subsection{Practical and Accurate Estimation of Demand and Resources}
\label{sec:estimation}

Both sides of Eq.~\ref{eq:e2e} must be estimated online on the proxy router, which serves thousands of requests per second; both estimators must therefore be \emph{practical}, cheap and free of engine internals, and \emph{accurate enough} for the routing decision, which needs the right feasibility set and the right ordering of instances rather than exact latencies.

\begin{figure}[t]
  \centering
  \begin{minipage}[t]{0.42\textwidth}
    \centering
    \includegraphics[width=0.9\linewidth]{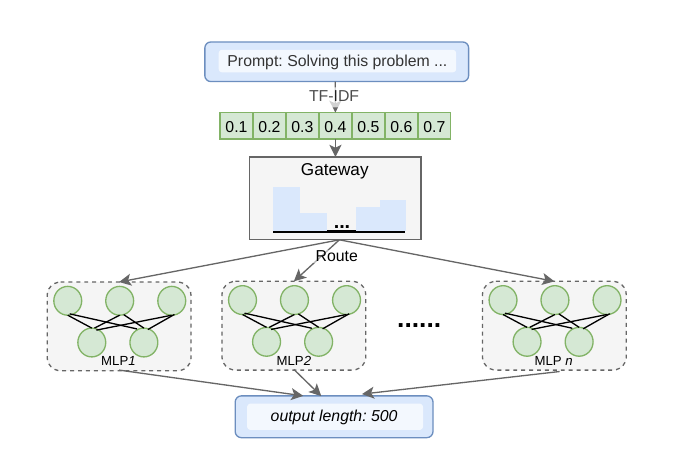}
    \caption{The output-length predictor: TF-IDF features, a gating router, and $K$ MLP experts.}
    \label{fig:output-length-predictor}
  \end{minipage}\hfill
  \begin{minipage}[t]{0.54\textwidth}
    \centering
    \includegraphics[width=0.9\linewidth]{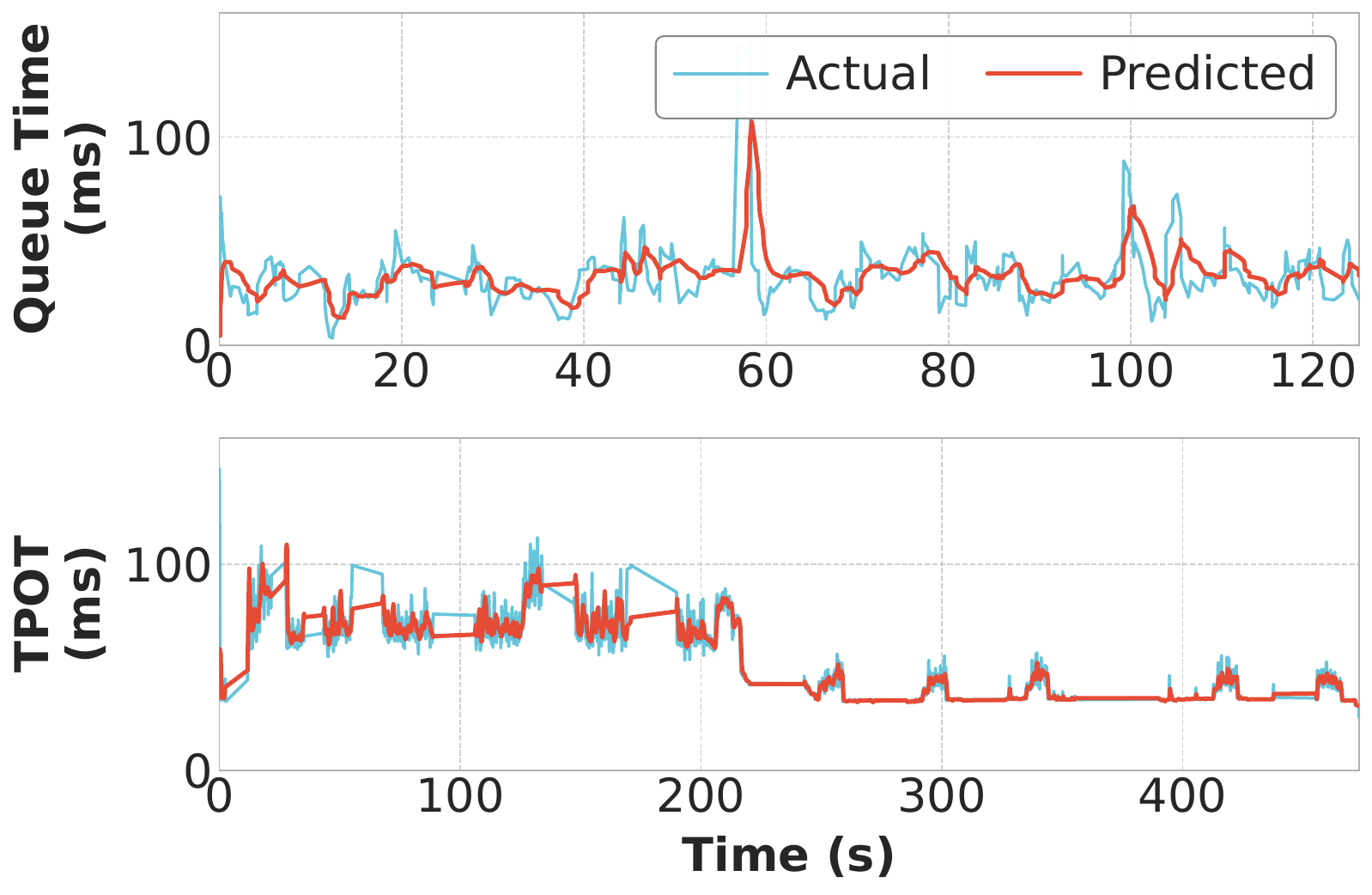}
    \caption{Estimated versus actual queuing time and TPOT on an A40 instance replaying the Mooncake trace.}
    \label{fig:queueing_time-and-tpot}
  \end{minipage}
\end{figure}

\phm{Demand: output-length prediction.}
Agentic inferences span a much wider output-length range than chatbot inferences because they span task types, and the task type is the dominant predictor of length: a database-operating request~\cite{phan2025askdb} is short, a code-generation request~\cite{liu2025sew} long.
This condition is visible in the prompt but must be handled \emph{implicitly}, without assuming a fixed task set.
Existing predictors either pay for a Transformer encoder, as $S^3$~\cite{jin2023s} does with a fine-tuned DistilBERT, or lose accuracy by ignoring the prompt, as history-based~\cite{gong2025past} and single-MLP~\cite{wang2025adaptive} predictors do.
\sys instead uses a mixture of small experts\footnote{Mixture-of-experts prediction has been used before, by MoPE in Equinox~\cite{wei2025equinox}, but for a fundamentally different job: MoPE predicts once, on arrival, to score fairness inside a single server, whereas our predictor exists to drive a global routing decision across heterogeneous instances and to be re-checked while the request runs, so it must be cheap enough for a proxy that handles thousands of requests per second, re-invocable on the generated prefix, and able to signal its own uncertainty. Technically, MoPE hard-routes each prompt to one of three fine-tuned BERT experts partitioned by output length; ours softly mixes nine lightweight MLP experts partitioned by joint input--output tiers through a learned gate, is re-invoked every $\tau$ iterations on the generated prefix, and exposes the dispersion of expert outputs as a confidence signal.} (Fig.~\ref{fig:output-length-predictor}): a feature vector $\mathbf{h}_r$ is obtained by TF-IDF vectorization~\cite{sparck1972statistical} of the token window so far, a two-layer gating router outputs $\mathbf{p}_r=\textit{softmax}(R(\mathbf{h}_r))\in\Delta^{K}$, and the prediction is the weighted combination $\hat{y}_r=\sum_{k=1}^{K} p_{r,k}\,E_k(\mathbf{h}_r)$ of $K$ four-layer MLP experts, 45.1M parameters in total.
Training proceeds in two phases: half of the data is partitioned into $K$ subsets by discretizing input and output lengths into $\sqrt{K}$ tiers each ($K=9$ in practice) and one expert is trained per subset; the router is then trained on the other half with the experts frozen, so that its gate learns to recognize the implicit task type.
TF-IDF is chosen for cost, not accuracy: replacing it by learned sentence embeddings with the same architecture changes the normalized error only from 1.00 to 0.98, while the fine-tuned DistilBERT predictor reaches 1.42 (Appendix~\ref{subsec:appendix_robustness}), and TF-IDF needs no encoder on the router, keeping prediction at 2.5\,ms per request.

\phm{Resources: instance-capability estimation.}
The resource coefficients $q_g$, $p_g$, and $d_g$ are hard to compute from configuration: under continuous batching in engines such as vLLM~\cite{vllm_github}, the speed a request experiences depends on the shapes of all requests sharing the batch, and it changes over the many decode steps of an agentic inference.
Two observations make estimation practical.
First, the routing decision needs a valid ordering of instances and a valid feasibility set, not exact latencies.
Second, an instance batches many requests per iteration and its configuration rarely changes, so by the law of large numbers~\cite{revesz2014laws} its per-iteration time is stable over short periods.
\sys therefore estimates $q_g$, $p_g$, and $d_g$ in a \emph{black-box} manner, from their recently observed values rather than from engine internals, and smooths them with an exponential moving average against transient jitter.
Fig.~\ref{fig:queueing_time-and-tpot} shows that the estimates track the actual queuing time and TPOT closely when replaying the Mooncake trace~\cite{qin2025mooncake} with mixed BIRD-bench~\cite{li2023can} and LiveCodeBench~\cite{jain2024livecodebench} requests on an A40 GPU.

\subsection{Globally Coordinated Routing with Runtime Rectification}
\label{sec:routing}

\phm{Just-enough selection.}
With the estimates in hand, \sys realizes the global principle with a rule that is simple to state and cheap to run.
Since Fig.~\ref{fig:gpu-letancy} shows that stronger GPUs are faster for every request, a greedy router would send every request to the strongest available one; but finishing a request earlier than its deadline adds nothing to goodput and takes capacity from the urgent requests that follow.
\sys therefore computes the feasible set $\mathcal{C}_r=\{g: T(r,g)\le D_r\}$ from Eq.~\ref{eq:e2e} and routes the request to the instance in $\mathcal{C}_r$ with the \emph{weakest} capability, i.e., the largest $d_g$, reserving stronger instances for requests that need them.
This is not an engineering shortcut but the per-request form of the global objective: each request consumes the least capable resource that keeps it within its deadline, which is precisely the locally-suboptimal, globally-optimal behavior that Sec.~\ref{sec:mot_challenges} found missing in existing routers, at a cost of $O(M)$ per request in an $M$-instance cluster.
When $\mathcal{C}_r$ is empty, no instance is predicted to meet the deadline and \sys rejects the request, so that capacity is not spent on work that cannot count toward goodput; Sec.~\ref{sec:deep-dive-eval} shows that this admission rule alone contributes 7.7\% goodput.
The complete procedure is given in Appendix~\ref{subsec:algorithm}.

\phm{Runtime rectification.}
Because the coefficients are estimates, a one-shot decision can turn out wrong: the output may run longer than predicted, or the chosen instance may slow down under new load.
After every $\tau$ inference iterations ($\tau=50$ by default), \sys re-predicts the remaining output length of each active request from its generated prefix and refreshes the instance estimates; if the expected finish time now exceeds the deadline, the proxy migrates the request to a stronger instance, again chosen by the just-enough rule.
Migration transfers the \emph{token IDs} rather than the KV cache: agentic inferences have long contexts and many decode iterations~\cite{wu2025combating}, so the KV cache is large while recomputing the prefill is cheap, and token-ID transfer trades a small prefill cost for a $7.1$--$15.3\times$ lower migration latency (Fig.~\ref{fig:micro-migration}), which is what makes rectification affordable at run time.

\section{Implementation}
\label{sec:implementation}

\begin{figure}[t]
        \centering
        \includegraphics[width=0.5\linewidth]{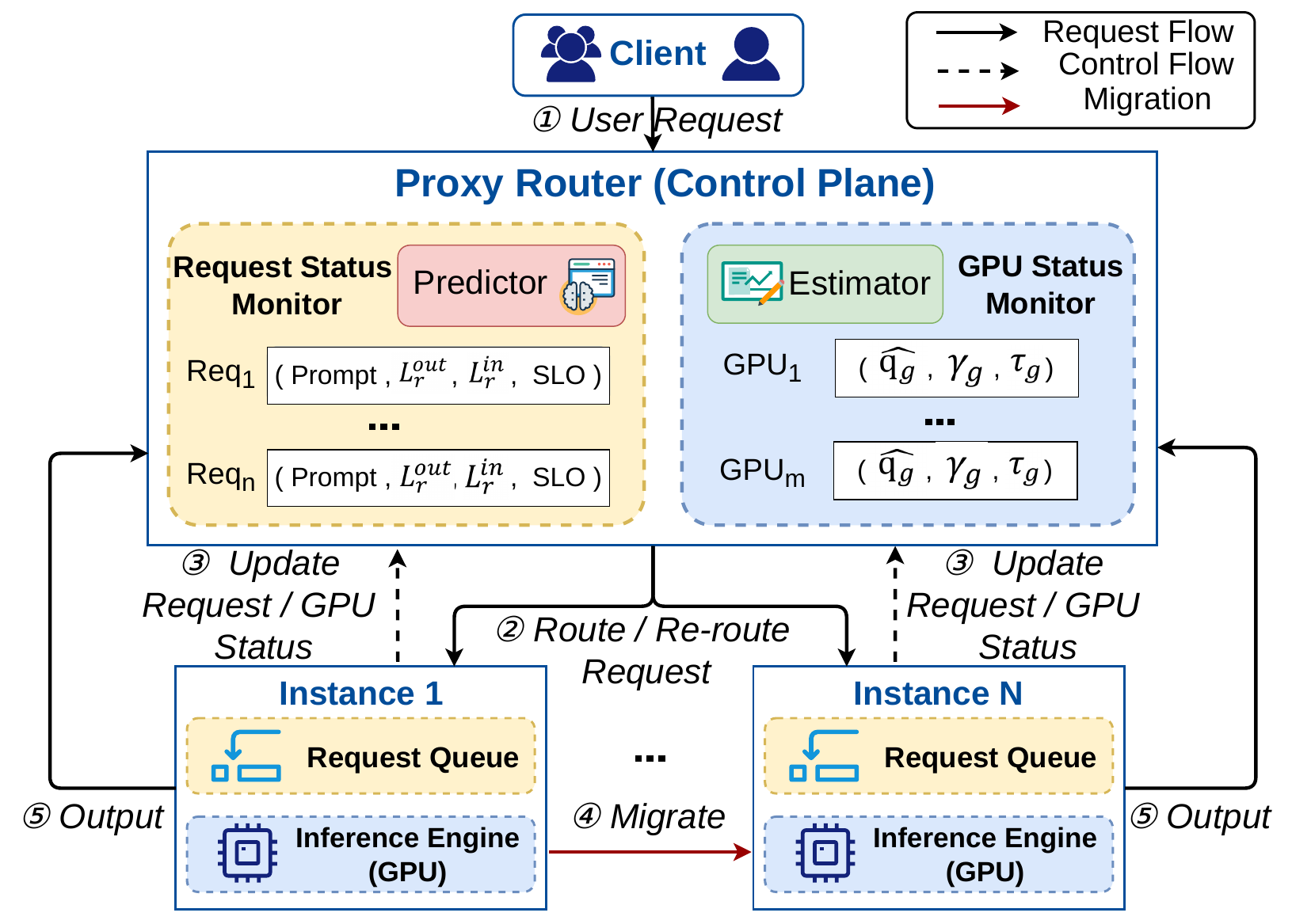}
        \caption{\sys architecture and workflow.}
        \label{fig:overview}
\end{figure}

Fig.~\ref{fig:overview} shows how the pieces fit into a serving stack.
A \texttt{RequestStatusMonitor} and a \texttt{GPUStatusMonitor} on the proxy router maintain the demand- and resource-side estimates of Sec.~\ref{sec:estimation}; the router applies the just-enough rule of Sec.~\ref{sec:routing} to each arriving request and issues migrations when the periodic risk check fires.
We implemented \sys in about 2.5k lines of Python on top of vLLM~\cite{vllm_github}; the source code is provided as supplementary material.
Each backend runs an unmodified inference engine except for a hook that re-predicts the remaining output length every $\tau$ generated tokens, with all such predictions batched on the backend, and reports SLO risk to the proxy through a \texttt{/check\_migration} interface; the proxy then re-evaluates feasibility, selects the target instance, and resumes generation there from the transferred token IDs.
The demand predictor is trained offline on 8,680 samples for 500 epochs, which takes 25 minutes, and is loaded once on the proxy and on each backend.
Because prediction is batched and routing is $O(M)$, the per-request overhead of the router stays at milliseconds even for hundreds of instances (Sec.~\ref{sec:deep-dive-eval}).

\section{Evaluation}
\label{sec:eval}

\begin{figure*}[t]
  \centering
  \begin{subfigure}{\textwidth}
    \centering
    \includegraphics[width=\textwidth]{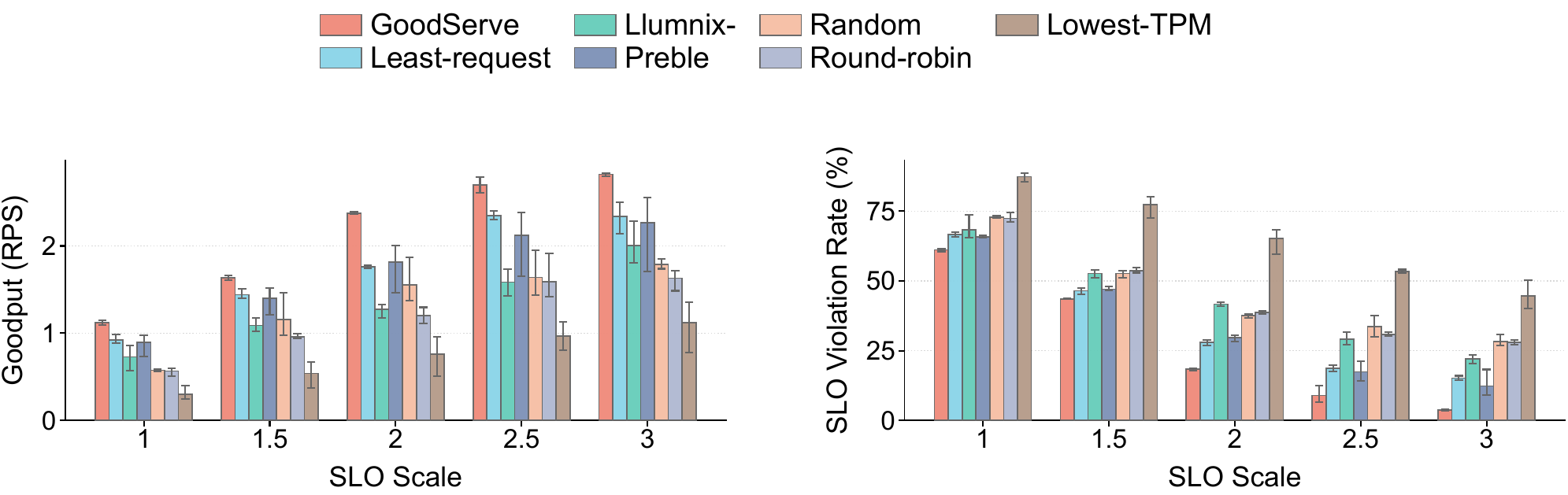}
  \end{subfigure}
  \\[\smallskipamount] 
  \small (a) Performance results on Llama3.1-8B-Instruct
  
  \vspace{0.4cm} 

  \begin{subfigure}{\textwidth}
    \centering
    \includegraphics[width=\textwidth]{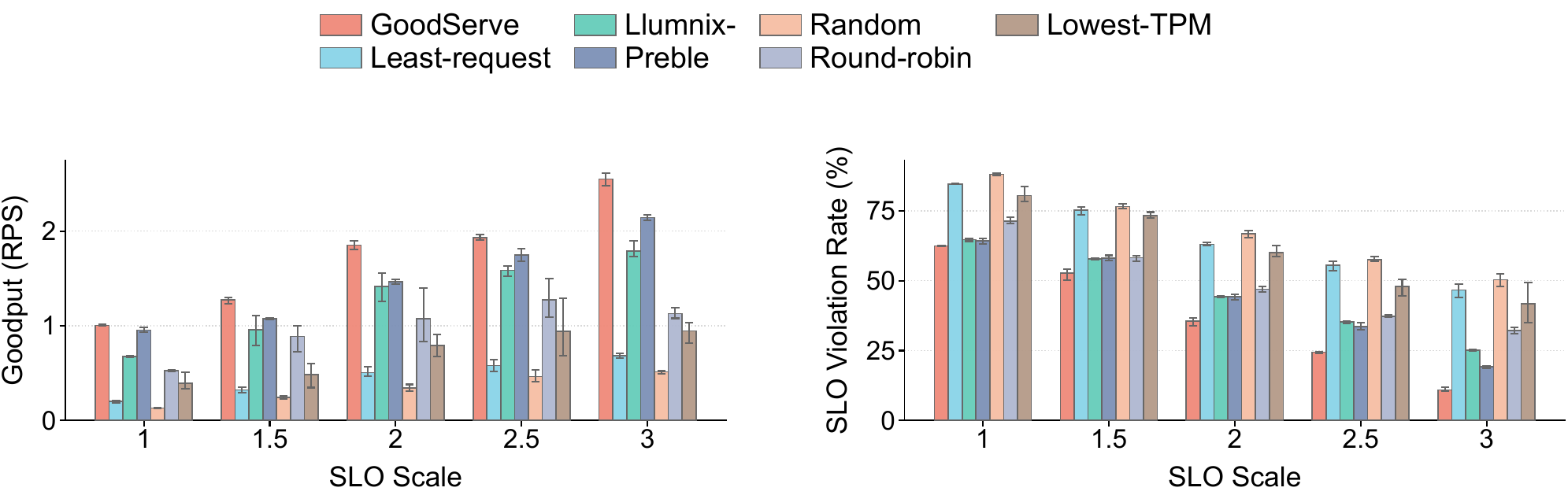}
  \end{subfigure}
  \\[\smallskipamount]
  \small (b) Performance results on Qwen2.5-14B-Instruct

  \vspace{0.3cm}
  \caption{End-to-end performance under different request routing methods.} 
  \label{fig:overall-comparison}
\end{figure*}

\subsection{Experimental Setup}
\label{sec:eval1}

\textbf{Hardware and models.}
Our experiments are conducted on a heterogeneous cluster of four instances, each equipped with distinct NVIDIA GPUs: H800 (80GB), A800 (80GB), A40 (48GB), and V100 (32GB, with a TP level of 2).
These instances are interconnected by a 10Gbps Ethernet network, and the proxy router is hosted on the A40 instance.
Four GPU generations is the scale at which heterogeneous serving is studied in the literature~\cite{griggs2024m}, and it is the degree of heterogeneity, not the raw GPU count, that routing exploits; routing overhead at cluster scale is measured separately with the real router over simulated backends (Sec.~\ref{sec:deep-dive-eval}).
We use two representative instruction-tuned models, Llama3.1-8B-Instruct \cite{grattafiori2024llama} and Qwen2.5-14B-Instruct \cite{qwen2025qwen25technicalreport}, as backend LLMs. 


\textbf{Workloads.}
We construct a mixed workload suite from three popular agentic benchmarks: BIRD-bench \cite{li2023can}, SWE-bench \cite{jimenez2023swe} and LiveCodeBench \cite{jain2024livecodebench}. 
We replay the Mooncake production traces~\cite{qin2025mooncake} to set the request submission time.
To set the E2E-SLOs, following established methodologies~\cite{li2023alpaserve, peng2025hexgen, zhang2025jitserve}, we first measure the median request execution time by running each request alone on a mid-tier GPU (NVIDIA A800), and then scale that base latency by multiple relaxation factors ($1\times$, $1.5\times$, $2\times$, $2.5\times$ and $3\times$) to emulate diverse SLO urgency levels.
The temperature is set to 0 to ensure output consistency across the deadline-setting stage and the testbed-measurement stage.
Prefix-cache hits are emulated by hashing prompts and skipping the prefill of cached blocks, which reproduces their effect on latency exactly; on this workload, 29\% of prompt tokens (107 of 460 on average) hit a cached prefix on the chosen instance.

\textbf{Baselines and Metrics.}
We compare \sys with diverse baselines described in Sec.~\ref{sec:mot_challenges}: the built-in \emph{random} and \emph{least-request} methods in AIBrix~\cite{team2025aibrix}, as well as the \emph{round-robin}~\cite{yu2025efficient}, \emph{lowest-TPM}~\cite{litellm2026}, \emph{Preble}~\cite{srivatsa2024preble} and \emph{Llumnix}~\cite{sun2024llumnix} methods.  
Meanwhile, we adopt two metrics: (1) \emph{goodput}, which is the average number of requests completing within their E2E-SLO \emph{per second}, and (2) \emph{SLO violation ratio}, which is the overall ratio of requests failing to satisfy the E2E-SLO requirements.

\subsection{End-to-End Performance}
\label{sec:macro_eval}

In Fig.~\ref{fig:overall-comparison}, we depict the end-to-end results across different SLO scales for both 8B and 14B models.
We repeat each experiment five times and report the max/min values as error bars.
As shown in Fig.~\ref{fig:overall-comparison}, \sys consistently delivers the best goodput and the lowest SLO violation rate. 
For example, for the 8B model at a medium SLO scale of 2, \sys surpasses the second-best (\emph{Preble}) by 31.0\% in goodput.  
Meanwhile, \sys generally provides larger benefits under looser SLO requirements. For the 14B model, its goodput improvement over the second-best baseline ranges from 5.2\% to 26.6\% across different SLO scales.
This is because a looser SLO requirement allows for more \emph{locally-suboptimal} yet \emph{globally-optimal} routing decisions, which existing SLO-unaware methods all fail to make. 

\subsection{Deep Dive Experiments}
\label{sec:deep-dive-eval}

In this part, we conduct a series of deep-dive experiments.
Unless otherwise specified, we use the 8B model at an SLO scale of 3. 

\begin{figure*}[t]
  \centering

  \begin{minipage}[t]{0.24\textwidth}
    \vspace{0pt}
    \centering
    \includegraphics[width=\linewidth,height=3.0cm,keepaspectratio]{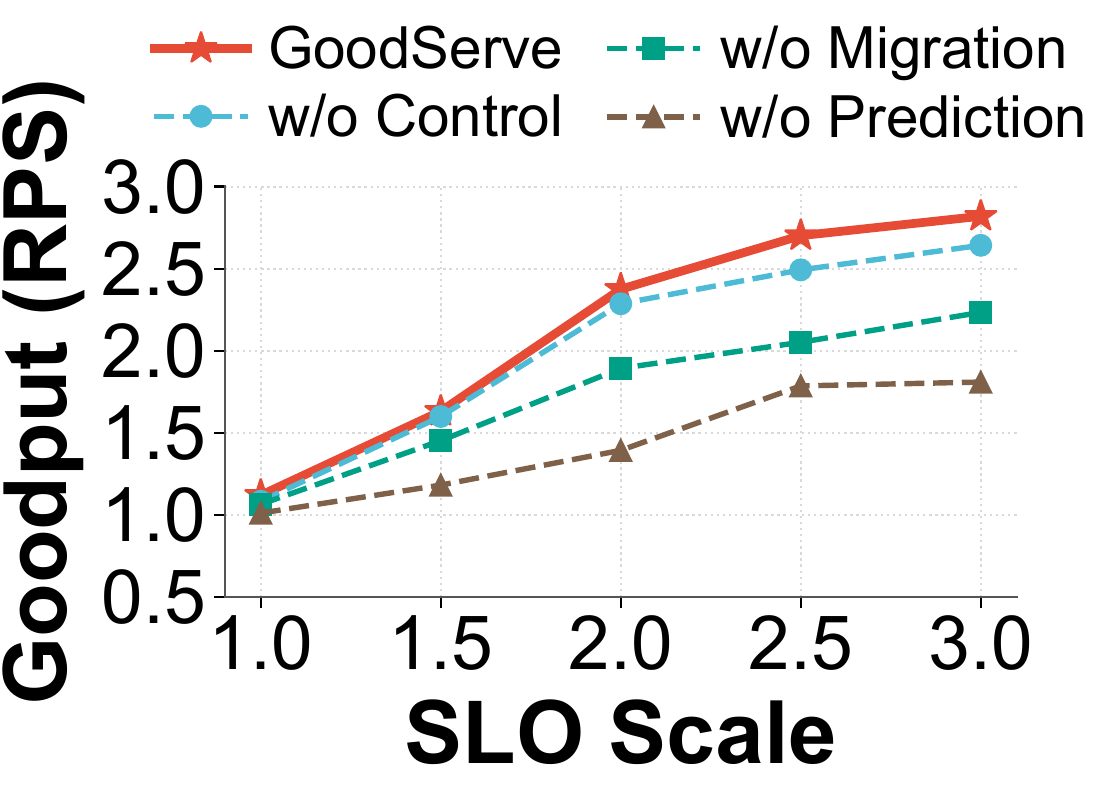}
    \captionof{figure}{Effectiveness of admission control, prediction, and migration innovations in \sys.} 
    \label{fig:ablation-8b}
  \end{minipage}
  \hfill
  \begin{minipage}[t]{0.48\textwidth}
    \vspace{0pt}
    \centering
    \begin{subfigure}[t]{0.48\linewidth}
      \centering
      \includegraphics[width=\linewidth,height=3.0cm,keepaspectratio]{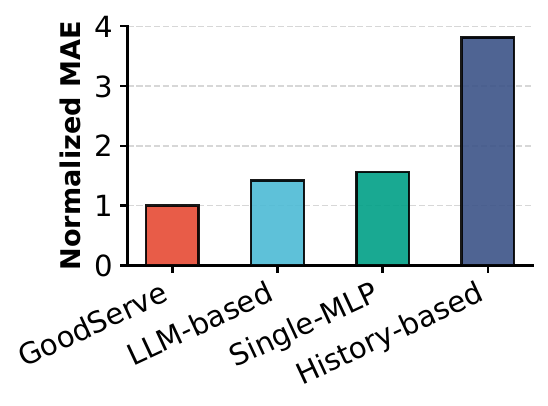}
      \vspace{-0.18in}
      \caption{Prediction error.}
      \label{fig:micro-pred-mae}
    \end{subfigure}
    \hfill
    \begin{subfigure}[t]{0.48\linewidth}
      \centering
      \includegraphics[width=\linewidth,height=3.0cm,keepaspectratio]{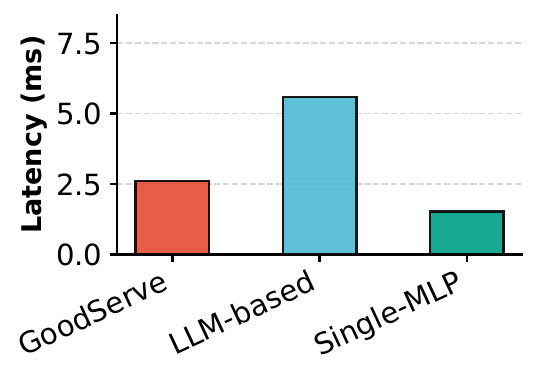}
      \vspace{-0.18in}
      \caption{Prediction latency.}
      \label{fig:micro-pred-latency}
    \end{subfigure}
    
    \captionof{figure}{Accuracy and overhead comparison among different prediction methods.} 
    \label{fig:microbench-prediction}
  \end{minipage}
  \hfill
  \begin{minipage}[t]{0.24\textwidth}
    \vspace{0pt}
    \centering
    \includegraphics[width=\linewidth,height=3.0cm,keepaspectratio]{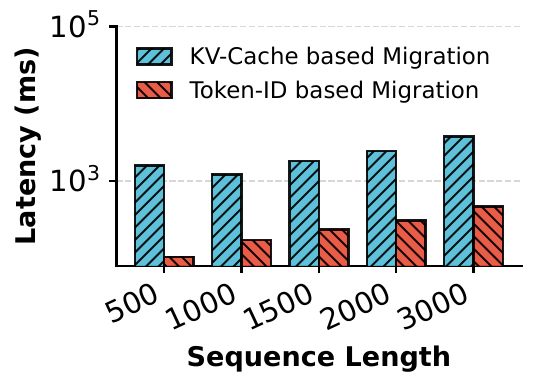}
    \captionof{figure}{Average migration latency under different state transferring methods.} 
    \label{fig:micro-migration}
  \end{minipage}

\end{figure*}

\textbf{Ablation study.}
Recall that \sys adopts admission control, an MoE-style request length predictor, and runtime migration to improve goodput and robustness. To evaluate their necessity, we create three variants: (1) \emph{\sys without admission control}, which admits all incoming requests; (2) \emph{\sys without prediction}, which replaces the original predictor with a history-based one~\cite{gong2025past}, as prediction itself cannot be disabled; and (3) \emph{\sys without migration}, which disables runtime request migration. As shown in Fig.~\ref{fig:ablation-8b}, all three components are indispensable.
For example, at an SLO scale of 2.5, removing the admission control, runtime migration, and MoE-style predictor reduces the goodput by 7.7\%, 24.0\%, and 33.9\%, respectively.
We note that their importance increases under looser SLOs, because a looser SLO allows for more flexible routing optimizations. 



\textbf{Superiority of our MoE-style demand predictor.}
We check the superiority of our MoE-style predictor over the other ones discussed in Sec.~\ref{sec:estimation}, in terms of both accuracy and overhead.
Fig.~\ref{fig:microbench-prediction} shows the normalized \emph{Mean Absolute Error} (MAE) for different predictors. 
It suggests that our MoE-style predictor adopted in \sys can yield the best accuracy---with an error reduction of $1.4\times$ compared to the LLM-based one~\cite{jin2023s}, and up to $3.8\times$ compared to the history-based one~\cite{gong2025past}.
Meanwhile, its prediction latency (around 2.5 ms per request) is lower than the LLM-based one. Furthermore, replacing our predictor with the LLM-based, Single-MLP, and history-based ones reduces the goodput by 12.5\%, 16.7\%, and 20.4\%, respectively. Moreover, even an Oracle predictor with exact output-length knowledge improves goodput by only 5.1\% over our predictor.

\textbf{Superiority of our Token-ID based request migration method.}
Recall that in Sec.~\ref{sec:routing}, for request migration we choose to transfer the Token-IDs instead of the KV cache.
To evaluate its superiority, we refer to Fig.~\ref{fig:micro-migration}, which shows the average migration latency for requests of different lengths. 
Compared with the KV-cache-based method, token-ID based migration achieves $7.1\times$ to $15.3\times$ lower latency. 

\captionsetup{
  font=small,
  labelfont=bf,
  justification=justified,
  singlelinecheck=false,
  skip=4pt
}
\captionsetup[subfigure]{
  font=small,
  justification=centering,
  singlelinecheck=false
}

\begin{figure*}[t]
  \centering
  \noindent
  \begin{minipage}[b]{0.50\textwidth}
    \centering
    \vspace{0pt}

    \captionsetup{type=figure}
    \phantomcaption
    \label{fig:sensitivity-trace}
    \setcounter{subfigure}{0}

    \begin{subfigure}[t]{0.46\linewidth}
      \centering
      \includegraphics[width=\textwidth]
      {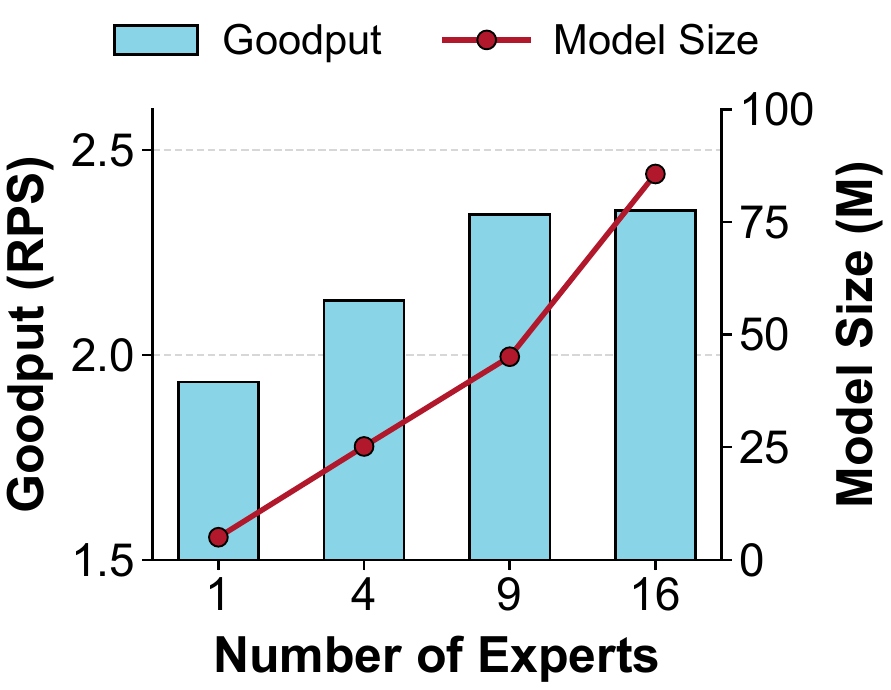}
      \caption{Number of experts.}
      \label{fig:sens-moe}
    \end{subfigure}%
    \hfill%
    \begin{subfigure}[t]{0.51\linewidth}
      \centering
      \includegraphics[width=\textwidth]
      {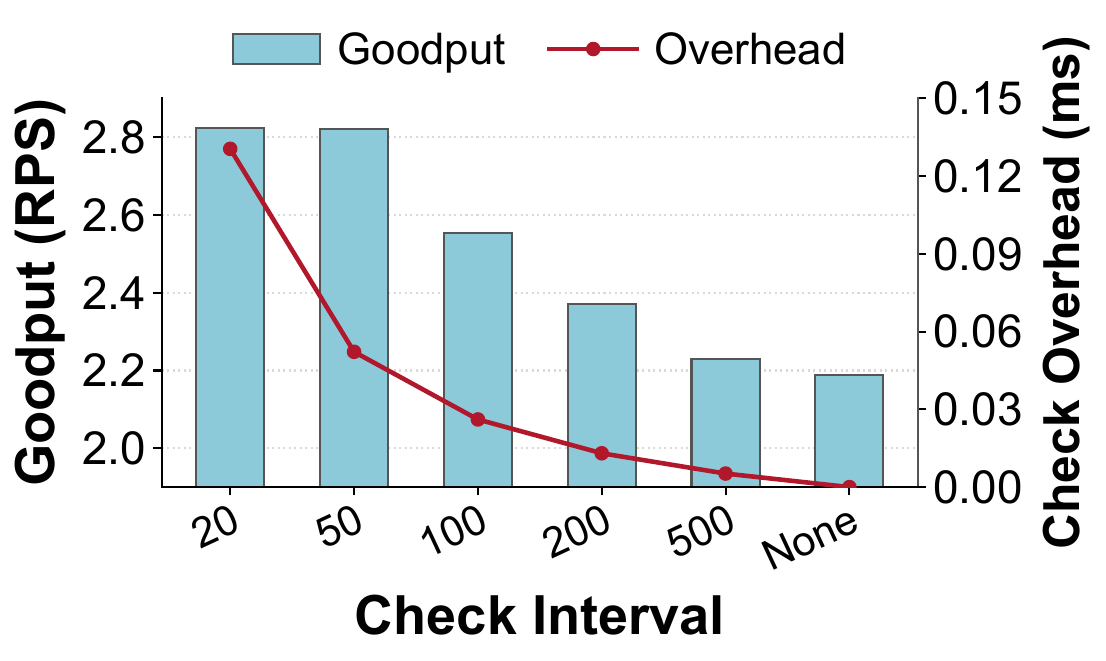}
      \caption{Status recheck interval.}
      \label{fig:sens-slo-recheck}
    \end{subfigure}

    \vspace{0.35em}

    \makebox[\linewidth][c]{%
      Figure~\thefigure: Hyper-parameter sensitivity analysis.%
    }
  \end{minipage}%
  \hfill%
  \begin{minipage}[b]{0.235\textwidth}
    \centering

    \vspace{1.5mm}

    \captionsetup{type=figure}
    \phantomcaption
    \label{fig:routing_overhead}

    \includegraphics[width=\textwidth]
    {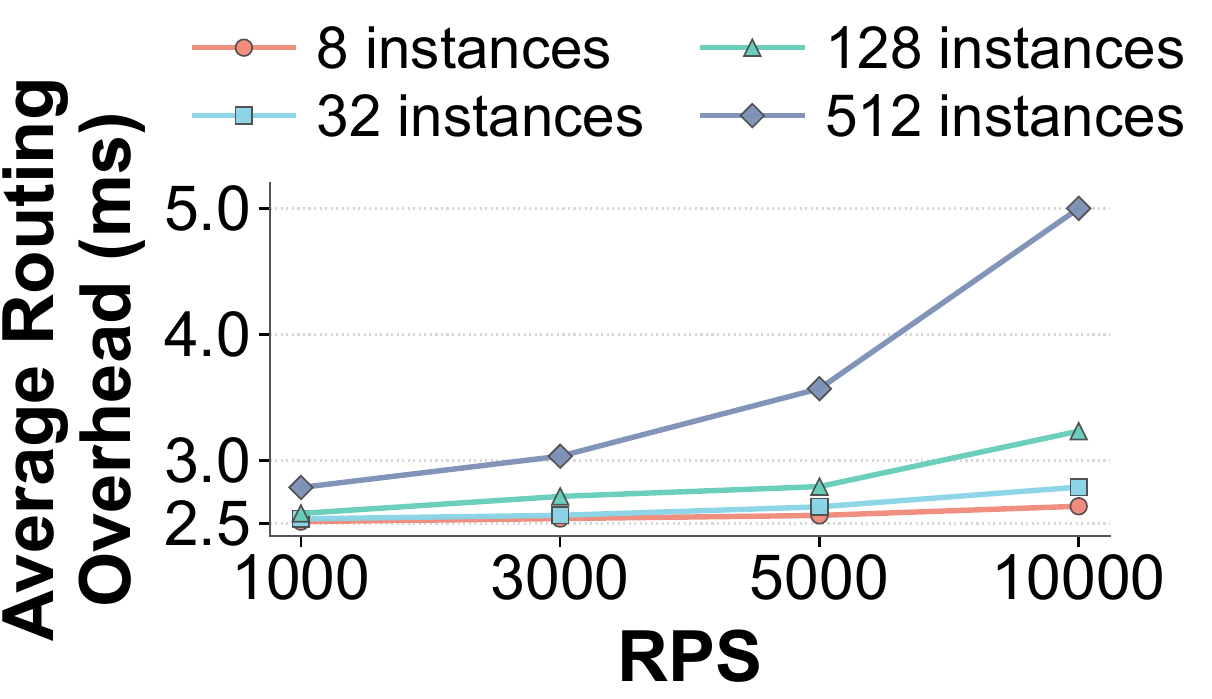}

    \vspace{0.35em}

    \caption*{%
      Figure~\thefigure:
      Routing overheads at varying cluster size and request intensity.%
    }
  \end{minipage}%
  \hfill%
  \begin{minipage}[b]{0.235\textwidth}
    \centering
    \vspace{0pt}

    \captionsetup{type=figure}
    \phantomcaption
    \label{fig:bursty-traffic}

    \includegraphics[width=\textwidth]
    {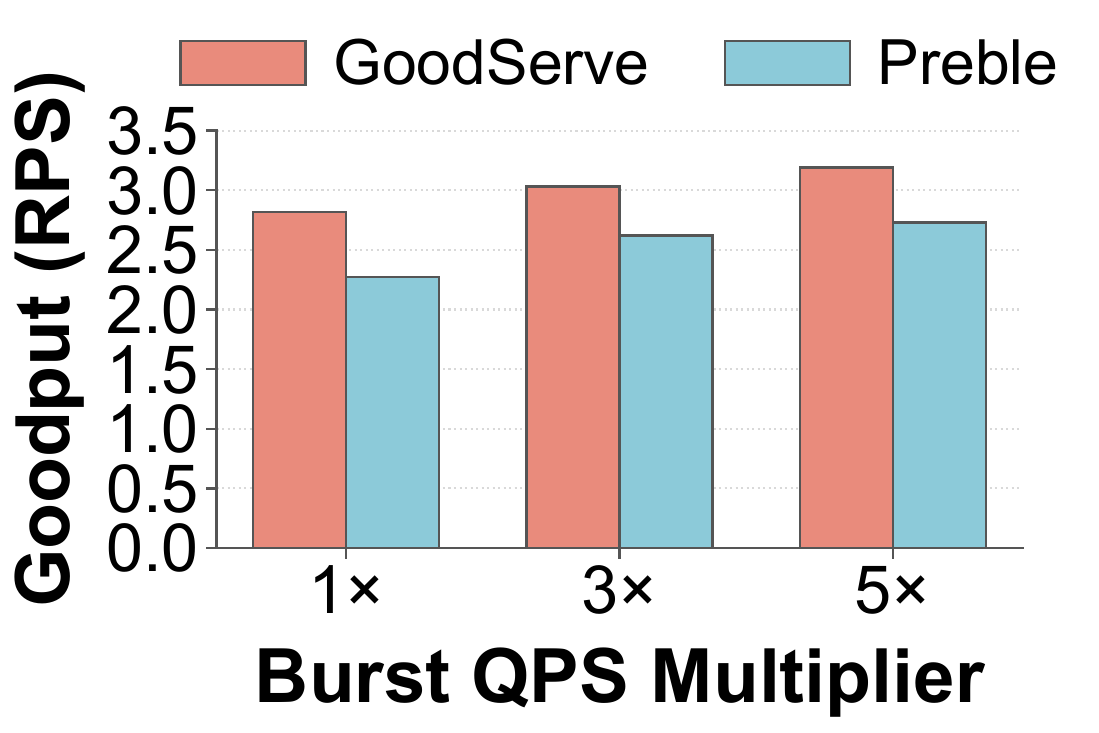}

    \vspace{0.35em}

    \caption*{%
      Figure~\thefigure:
      Goodput under varying burst traffic intensity.%
    }
  \end{minipage}
\end{figure*}

\textbf{Hyper-parameter sensitivity analysis.}
\sys introduces two hyper-parameters: the number of experts $K$ in the predictor (Sec.~\ref{sec:estimation}), and the status checking frequency $\tau$ (Sec.~\ref{sec:routing}). 
Here we evaluate \sys's performance by varying their values.
As shown in Fig.~\ref{fig:sens-moe}, while adopting more experts can yield higher accuracy and consequently higher goodput, $K=9$ can already yield sufficiently good performance, much better than $K=4$ yet comparable to $K=16$.
Meanwhile, regarding $\tau$, Fig.~\ref{fig:sens-slo-recheck} suggests that a higher checking frequency does improve goodput.

\textbf{Overhead analysis with large-scale simulation.}
In production clusters, there may be hundreds of instances. 
Since our core innovation lies in the proxy router instead of on the serving engine, to check \sys's scalability we resort to large-scale simulations.
Specifically, we configure a set of virtual IPs each corresponding to a simulated local inference engine. 
We respectively simulate 8, 32, 128 and 512 instances, and for each case we vary the RPS from 1000 to 10000---all requests handled by a single router.
As shown in Fig.~\ref{fig:routing_overhead}, thanks to the lightweight prediction model (Sec.~\ref{sec:estimation}), the simple heuristic (Sec.~\ref{sec:routing}) as well as the implementation optimizations (distributed and batched predictor execution), the routing overhead of \sys is indeed quite small: the per-request routing latency is still marginal (5.0 ms) even when serving highly-intense workloads (RPS=10000) over 512 nodes.

\textbf{Bursty traffic.}
To evaluate scheduling under bursty arrivals, we inject a controlled local burst into the original trace by selecting a contiguous request segment and compressing its inter-arrival times to increase the local arrival rate to $1\times$, $3\times$, and $5\times$ the original rate, while keeping the remaining trace unchanged. As shown in Figure~\ref{fig:bursty-traffic}, GoodServe consistently outperforms Preble across all burst intensities, achieving at least 15.6\% higher goodput.
Appendix~\ref{subsec:appendix_robustness} further reports \sys's goodput under out-of-distribution tasks, non-zero decoding temperature, a predictor transferred from a different model, an SLO-aware routing baseline distilled from JITServe, and semantic-embedding features for the predictor.

\section{Related Work and Discussion}
\label{sec:discussion}

\phm{Positioning against heterogeneous and SLO-aware serving systems.}
Several systems touch heterogeneous or SLO-aware serving, but each decides a different variable.
Helix~\cite{mei2025helix} decides how one model's layers are placed across heterogeneous nodes and which node path a request traverses, targeting aggregate throughput; M\'elange~\cite{griggs2024m}, MILP-based provisioning studies~\cite{jiang2025demystifying}, SageServe~\cite{jaiswal2025serving}, and \textit{llm-d}~\cite{llmd2026wva} decide the GPU composition, deployment configuration, or scale of a deployment; JITServe~\cite{zhang2025jitserve} schedules requests \emph{within} an instance under imprecise request information and does not route across heterogeneous backends; Llumnix~\cite{sun2024llumnix} and Preble~\cite{srivatsa2024preble} route across instances but without deadline awareness, and Llumnix migrates KV cache rather than token IDs.
\sys takes the instance set as given and decides, per request and per deadline, which self-contained instance serves it; it can run standalone or as the routing layer beneath a provisioning system, and it is, to our knowledge, the first routing work for agentic inferences whose primary metric is TTLT.
An SLO-aware routing baseline distilled from JITServe's earliest-available-slot principle reaches 1.79 goodput versus 2.68 for \sys in Appendix~\ref{subsec:appendix_robustness}, because it remains greedy in its instance choice.

\phm{Applicability.}
For inferences without E2E-SLOs, \sys can fall back to conventional routing without affecting its other components, and routing relaxed agentic inferences to weaker instances in fact improves the TTFT and TPOT of co-served chatbot inferences.
In homogeneous clusters the instances still differ in queuing time and batch size at any instant, so the same rule applies.
For PD-disaggregated serving~\cite{qin2025mooncake}, Eq.~\ref{eq:e2e} can model the prefill and decode terms on the prefill-only and decode-only instances of each candidate pair, and the just-enough rule carries over.

\section{Conclusion}
\label{sec:conlucsion}

Agentic LLM inferences are judged by whether their whole output arrives before a deadline, and serving them on heterogeneous GPUs turns routing into a global goodput problem that greedy, locally optimal routers cannot solve.
\sys formulates that problem with a latency model coupling request demand and instance capability, and solves it in a predict-and-rectify manner: practical estimators for both sides, a just-enough rule that realizes globally coordinated routing at $O(M)$ per request, and token-ID migration that corrects estimation errors at run time.
On a heterogeneous testbed it raises goodput by up to 31.0\% over the best existing router, and its routing overhead remains negligible even at the scale of hundreds of instances.



\bibliography{iclr2027_conference}
\bibliographystyle{iclr2027_conference}

\appendix

\newpage
\clearpage
\appendix

\section{Appendix}

\subsection{Notations used in \sys}
\label{subsec:notion}

\begin{table}[H]
\centering
\small
\setlength{\tabcolsep}{6pt}
\begin{tabularx}{\columnwidth}{lX}
\toprule
\textbf{Notation} & \textbf{Description} \\
\midrule
$r$, $g$ & Request index, GPU instance index \\
$\mathcal{R}$, $\mathcal{G}$ & Set of requests, set of GPU instances \\
$D_r$ & End-to-end latency deadline (E2E-SLO) of request $r$ \\
$L^{\mathrm{in}}_r$, $L^{\mathrm{out}}_r$ & Input length and (predicted) output length of request $r$ \\
$H_{r,g}$ & Reusable (cache-hit) prefix length of request $r$ on GPU $g$ \\
$q_g$ & Estimated average queuing delay on GPU $g$ \\
$p_g$, $d_g$ & Estimated average per-token prefill and decode latencies on GPU $g$ \\
$x_{r,g}$ & Binary routing variable: $1$ if request $r$ is routed to GPU $g$ \\
$T(r,g)$ & Predicted end-to-end latency of request $r$ if routed to GPU $g$ (Eq.~\ref{eq:e2e}) \\
$\mathcal{C}_r$ & Feasible instance set of request $r$, $\{g : T(r,g)\le D_r\}$ \\
$\tau$ & Status re-check interval (inference iterations) \\
$\mathbb{I}[\cdot]$ & Indicator function \\
\bottomrule
\end{tabularx}
\caption{Key notations used in the routing formulation of \sys.}
\label{tab:overview_notation}
\end{table}

\subsection{Routing Algorithm}
\label{subsec:algorithm}

\begin{algorithm}[H]
\caption{Just-enough instance selection for an arriving request $r$}
\label{alg:backend_selection}
\begin{algorithmic}[1]
\State $\mathbf{h}_r \leftarrow \textsc{BuildFeatures}(r)$; \quad $L^{\mathrm{out}}_r \leftarrow \mathcal{P}(\mathbf{h}_r)$ \Comment{demand prediction}
\State $\mathcal{C}_r \leftarrow \emptyset$
\ForAll{$g \in \mathcal{G}$}
    \State $H_{r,g} \leftarrow \textsc{ReusePrefix}(r,g)$; \quad $(q_g,p_g,d_g) \leftarrow \textsc{InstanceEstimate}(g)$ \Comment{resource estimation}
    \State $T(r,g) \leftarrow q_g + p_g\,(L^{\mathrm{in}}_r - H_{r,g}) + d_g\,L^{\mathrm{out}}_r$
    \If{$T(r,g) \le D_r$ \textbf{and} GPU-$g$ has memory for $r$}
        \State add $g$ to $\mathcal{C}_r$
    \EndIf
\EndFor
\If{$\mathcal{C}_r = \emptyset$} \State \Return $\textsc{Reject}(r)$ \EndIf
\State \Return $g^\star \leftarrow \arg\max_{g\in\mathcal{C}_r} d_g$ \Comment{weakest feasible instance}
\end{algorithmic}
\end{algorithm}

Algorithm~\ref{alg:backend_selection} is invoked once when a request arrives and again whenever the periodic risk check of Sec.~\ref{sec:routing} finds that an active request is no longer expected to meet its deadline on its current instance; in the latter case the request's remaining output length replaces $L^{\mathrm{out}}_r$, its current instance is excluded from $\mathcal{G}$, and the selected instance becomes the migration target.

\subsection{Additional Robustness Experiments}
\label{subsec:appendix_robustness}

Unless noted otherwise, the experiments below use Llama-3.1-8B-Instruct at an SLO scale of 3 on the testbed of Sec.~\ref{sec:eval1}, and ``best baseline'' denotes the best-performing baseline of Sec.~\ref{sec:macro_eval} in each setting.

\paragraph{Prediction accuracy versus the routing rule.}
With the just-enough rule held fixed, we replace the predictor by an oracle that knows the true output length and by the history-based baseline predictor~\cite{gong2025past}.
\begin{center}\small
\begin{tabular}{@{}lc@{}}
\toprule
Configuration (routing identical) & Goodput (RPS) \\
\midrule
Just-enough routing + oracle output length & 2.86 \\
Just-enough routing + \sys predictor & 2.68 \\
Just-enough routing + history-based predictor & 2.23 \\
\bottomrule
\end{tabular}
\end{center}
Perfect prediction adds 6.7\% over our predictor, whereas the baseline predictor loses 16.8\%; the goodput that survives even the weakest predictor is attributable to the routing rule itself, and our predictor closes most of the remaining gap.

\paragraph{An SLO-aware routing baseline.}
JITServe~\cite{zhang2025jitserve} schedules within an instance and exposes no inter-instance routing policy.
We distill the routing analogue of its principle, earliest-available-slot-first, and compare it end to end; this construction is ours and should be read as a comparison against the principle rather than the JITServe system.
It reaches a goodput of 1.79 versus 2.68 for \sys: earliest-available-slot remains greedy in its instance choice and still occupies the most capable instances with requests that weaker instances could have completed in time.

\paragraph{Out-of-distribution tasks, decoding temperature, and predictor transfer.}
\begin{center}\small
\begin{tabular}{@{}lcc@{}}
\toprule
Setting & \sys & Best baseline \\
\midrule
Default workload & 2.68 & 2.27 \\
Workload extended with GSM8K requests (one third of the original size) & 2.63 & 2.25 \\
Decoding temperature 0.7 & 2.67 & 2.38 \\
Predictor trained for Qwen2.5-14B, serving Llama-3.1-8B & 2.59 & --- \\
\bottomrule
\end{tabular}
\end{center}
The advantage holds under task drift, sampling-based decoding, and a predictor transferred from a different model, the harshest of the three shifts; runtime rectification absorbs the added prediction error.

\paragraph{Semantic features for the predictor.}
Replacing TF-IDF features by LaBSE sentence embeddings, with the predictor architecture unchanged, gives a normalized MAE of 0.98 versus 1.00, while the fine-tuned DistilBERT predictor $S^3$~\cite{jin2023s} gives 1.42.
Learned embeddings add no accuracy for this task, where output length is governed by lexical and structural cues of the task type, and they require a 3\,GB encoder on the router; \sys uses TF-IDF for cost, not for accuracy.

\paragraph{Migrations under bursty arrivals.}
Under the $5\times$ burst intensity of Fig.~\ref{fig:bursty-traffic}, the number of migrations rises to $1.15\times$ that under normal traffic, which is the rectification mechanism absorbing the burst; the relative advantage of \sys stays above 15\% at every burst level.

\end{document}